\documentclass{icrc2009}

\usepackage{graphicx}   
\usepackage[caption=false]{caption}    
\usepackage[font=footnotesize]{subfig} 
\usepackage{fixltx2e}
\usepackage{url}
\usepackage{bm}

\newcommand{\shorttitle}[1]%
{\markboth{Proceedings of the 31\MakeLowercase{$^{st}$} ICRC, {\L}\'{o}d\'{z} 2009}{#1} }
\newcommand{\etal}{\MakeLowercase{\textit{et al. }}} 

\def\simleq{\; \raise0.3ex\hbox{$<$\kern-0.75em \raise-1.1ex\hbox{$\sim$}}\; }
\def\simgeq{\; \raise0.3ex\hbox{$>$\kern-0.75em \raise-1.1ex\hbox{$\sim$}}\; }

\newcommand{\MeV}{{\rm MeV}}
\newcommand{\GeV}{{\rm GeV}}
\newcommand{\TeV}{{\rm TeV}}

\newcommand{\kpc}{{\rm kpc}}

\newcommand{\cm}{{\rm cm}}

\newcommand{\s}{{\rm s}}

\newcommand{\pbar}{\bar{p}}

\usepackage{amsmath}

\begin{document}
  \title{A Combined Interpretation of Cosmic Ray Nuclei and Antiproton High Energy Measurements}
  
  \author{\IEEEauthorblockN{Carmelo Evoli\IEEEauthorrefmark{1},
      Daniele Gaggero\IEEEauthorrefmark{2}\IEEEauthorrefmark{3},
      Dario Grasso\IEEEauthorrefmark{2} and
      Luca Maccione\IEEEauthorrefmark{4}}
    \\
    \IEEEauthorblockA{\IEEEauthorrefmark{1}SISSA/International School for Advanced Studies, via Beirut, 2-4, I-34014 Trieste, Italy.}
    \IEEEauthorblockA{\IEEEauthorrefmark{2}INFN, Sezione di Pisa, Largo B. Pontecorvo, 3, I-56127 Pisa, Italy.}
    \IEEEauthorblockA{\IEEEauthorrefmark{3}Dipartimento di Fisica ``E. Fermi", Universit\`a di Pisa, Largo B. Pontecorvo, 3, I-56127 Pisa, Italy.}
    \IEEEauthorblockA{\IEEEauthorrefmark{4}DESY, Theory Group, Notkestrasse 85, D-22607 Hamburg, Germany}}
  
  \shorttitle{C. Evoli \etal - Combined Interpretation of CR Nuclei and \lowercase{$\pbar$}.}
  \maketitle
  
  \begin{abstract}
    In the last months several ballon and satellite experiments improved significantly our 
    knowledge of cosmic ray (CR) spectra at high energy. In particular CREAM allowed to 
    measure B/C, C/O and N/O ratios up to 1 TeV/n and PAMELA the $\bm{{\bar p}/p}$ ratio up to 100 GeV 
    with unprecedented accuracy. These measurements offer a valuable probe of CR propagation 
    properties. We performed a statistical analysis to test the compatibility of these 
    results, as well as other most significant experimental data, with the predictions of a 
    new numerical CR diffusion package (DRAGON). We found that above 1 GeV/n all data are consistent 
    with a plain diffusion scenario and point to well defined ranges for the normalization and energy 
    dependence of the diffusion coefficient.
  \end{abstract}
  
  
  \section{Introduction}
  
  The problems of origin and propagation of Cosmic Rays (CR) in the Galaxy are long standing questions which need
  the combination of several different observations in a wide energy range to be answered \cite{Strong:2007nh}. 
  
  The most realistic description of CR propagation is given by diffusion models. Two main approaches have been 
  developed so far: (semi-)analytical diffusion models (see e.g.~\cite{Berezinsky:book} and ref.s therein), which 
  solve the CR transport equation by assuming simplified distributions for the sources and the interstellar gas,
  and fully numerical diffusion models. Well known realizations of those two approaches are respectively the 
  {\it two-zone model} \cite{Maurin:01,Maurin:02} and the GALPROP \cite{Strong:98,Strong:04,GALPROPweb} and DRAGON 
  \cite{Evoli:2008dv} codes.
  Generally these models involve a large number of parameters which need to be fixed against the observations.  
  Their knowledge is crucial not only for CR physics but also to be able to constrain/determine the properties of 
  dark matter from indirect measurements. However, in spite of the strong efforts made on both observational and 
  theoretical sides most of those parameters are still poorly known. One of the reasons lies in the fact that best 
  quality data on CR spectra were available mainly at low energy ($\!\simleq 10~\GeV$/n). At these energies several 
  competing physical processes (e.g.~solar modulation, convection, re-acceleration) are expected to affect 
  significantly the CR spectra to an {\it a priori} unknown amount. Furthermore, at those energies uncertainties in 
  spallation cross section determinations are still sizeable. 
  At higher energies, however, only spatial diffusion and spallation losses are expected to shape the CR spectra, 
  the latter effect becoming less relevant with increasing energy. Hence, the study of high energy CR spectra could 
  allow to constrain the properties of the diffusion coefficient of CR in the Galaxy.

  The scarcity of observational data has precluded this possibility for long time. The situation improved recently as 
  the CREAM balloon experiment \cite{CREAM} measured the relative abundances of elements from boron to oxygen, and 
  especially the boron to carbon ratio (B/C), up to energies around $1~\TeV$/n.
  Furthermore, valuable complementary data were recently provided by the PAMELA satellite experiment \cite{PAMELA} 
  which measured the antiproton/proton ($\pbar/p$) ratio up to $100~\GeV$ with unprecedented accuracy. 
  Also antiprotons are expected to be produced by the spallation of primary CRs (mainly protons and Helium nuclei). 
  They provide, therefore, an independent test of the validity of CR propagation models \cite{Donato:2001,Moskalenko:2005} 
  and, once these are validated, a valuable probe of dark matter models (see e.g. \cite{Donato:2009}). 

  In this contribution we show that the data recently released by the CREAM and the PAMELA experiments can fit into 
  a unique plain diffusion (PD) CR propagation model. We report the main results of a statistical analysis aimed at 
  constraining the normalization and energy dependence of the diffusion coefficient of CRs in the Galaxy. 
  In order to check the possible dependence on low energy effects of these constraints, we also study as they change
  by varying the minimum energy $E_{\rm min}$ above which data are considered.

  \section{Model}
  \label{sec:code}
  
  In order to interpret the experimental data we need to adopt a theoretical model describing the propagation of CR 
  nuclei in the Galaxy between 1 and $10^{3}$ GeV/n. 
  At these energies, the propagation of stable CRs is known to obey the transport equation \cite{Ginzburg:64}
  \setlength{\arraycolsep}{0.0em}
  \begin{multline}
    \label{eq:transport}
    \frac{\partial N_i}{\partial t} = -{\nabla}\cdot \left( {\bf D}\cdot\,{\nabla}N_i \right)  + Q_{i}(E_k) + \\
    -c\,\beta\,n_{\rm gas}\,\sigma_{\rm in}(E_{k})N_{i} + \sum_{j>i}c\,\beta\, n_{\rm gas}\, \sigma_{ji}N_{j}\;,
  \end{multline}
  \setlength{\arraycolsep}{5pt}
  where $E_{k} \equiv (E-m_{A})/A$ ($E$ is the total energy of a nucleus with mass $m_{A} \simeq A\times m_{\rm pr}$) 
  is the kinetic energy per nucleon, constant during propagation as practically conserved in fragmentation reactions, 
  $\beta$ is the velocity of the nucleus in units of the speed of light $c$,  $\sigma_i$ is the total inelastic cross 
  section onto the ISM gas with density $n_{\rm gas}(r,z) $ and $\sigma_{ij}$ is the production cross-section of a 
  nuclear species $j$ by the fragmentation of the $i$-th one. We start the spallation routine from $A = 64$. 
  We disregard continuos energy losses, re-acceleration and convection, but we check {\it a posteriori} the validity 
  of this approximation against the experimental data. 

  We solve Eq.~(\ref{eq:transport}) in the stationary limit $\partial N_{i}/\partial t = 0$ adopting our numerical code 
  DRAGON \cite{Evoli:2008dv}. DRAGON was validated against well known public codes (GALPROP) and against experimental 
  data of secondary/primary ratios, as well as $\gamma$-ray data.

  We recall below the main assumptions we make.

  \subsection{Spatial diffusion}
  
  We assume cylindrical symmetry and that the regular magnetic field is azimuthally oriented 
  $({\bf B_0} = B_\phi(r,z)\,\hat{\bf{\phi}})$. 
  Under these conditions CR diffusion out of the Galaxy takes place only perpendicularly to ${\bf B_0}$. Therefore $D$ 
  represents in fact the perpendicular diffusion coefficient $D_\perp$. The dependence of $D$ on the particle rigidity 
  $\rho$ is (see e.g.~\cite{Ptuskin:97})
  
  \setlength{\arraycolsep}{0.0em}
  \begin{equation}
    \label{eq:diff_coeff}
    D(\rho, r,z) = D_0~\beta \left(\frac{\rho}{\rho_0}\right)^\delta\ ~ \exp\left\{|z|/z_t \right\}\;.
  \end{equation}
  \setlength{\arraycolsep}{5pt}

  \subsection{CR sources} 
  
  For the source term we assume the general form 
  \begin{equation}
    Q_{i}(E_{k},r,z) =  f_S(r,z)\  q^{i}_{0}\ \left(\frac{\rho}{\rho_0}\right)^{- \alpha_i} \;,
  \end{equation}
  imposing $Q_{i}(E_{k},r_{\odot},z_{\odot}) = 1$.

  We assume that the CR source spatial distribution $f_S(r,z)$ trace that of Galactic supernova remnants (SNRs)
  as modeled in \cite{Ferriere:01} on the basis of pulsar and progenitor star surveys \cite{Evoli:2007iy}. 

  The injection abundances $q^i_0$ are tuned so that the propagated spectra of primary and secondary species 
  match the observed ones.  

  For each value of $\delta$ in Eq.~(\ref{eq:diff_coeff}) $\alpha_i$ is fixed by the requirement that at high 
  energy $E_k \gg 100~\GeV$/n, at which spallation processes are almost irrelevant, the equality 
  $\alpha + \delta = 2.7$ is satisfied\footnote{In this regime, the theoretical expectation for the observed flux 
    $\Phi$ on Earth is $\Phi(E) \approx Q(E)/D(E) \sim E^{-(\alpha+\delta)}$ \cite{Berezinsky:book}.}.

  \section{Analysis and results}
  \label{sec:analysis}
  
  Since our main goal is to understand in particular the diffusion properties, we want derive constraints on $D_{0}$, 
  $\delta$ and $z_{t}$ in Eq.~(\ref{eq:diff_coeff}). 
  We consider these observables: N/O, C/O, B/C and $\bar{p}/p$ ratios. They are primary/primary and secondary/primary ratios. 
  The spectrum of secondary/primary ratios allows us to infer information directly on $\delta$ \cite{Berezinsky:book}, but 
  not separately on $D_{0}$ and $z_{t}$ (these observables are sensitive to the ratio $D_{0}/z_{t}$ \cite{Evoli:2008dv}). 
  A way to break this degeneracy is to consider unstable to stable ratios (e.g.~$^{10}$Be/$^{9}$Be), which are known to probe 
  the vertical height of the Galaxy \cite{Berezinsky:book}. In agreement with \cite{Evoli:2008dv,Moskalenko:2001qm} we infer 
  that $z_{t}$ should lie between 3 and 5 kpc.
  We account for a solar modulation potential $\Phi = 550~\MeV$ in the ``force-free" approximation \cite{Gleeson&Axford}.
  
  \subsection{Strategy}
  
  \subsubsection{B/C ratio}
  
  Once the spatial distributions of the CR sources and the ISM gas have been chosen, the  main parameters determining 
  the B/C in a PD model are the C/O and N/O injection ratios and the quantities $\delta$ and $D_0/z_t$ (which will be 
  always expressed in units of $10^{28}~\cm^2~\s^{-1}~\kpc^{-1}$ in this work) in Eq.~(\ref{eq:diff_coeff}). 

  We fix the source abundances of the oxygen and of primaries heavier than oxygen by requiring that they match the observed 
  abundances in CRs at $E \sim 1-10~\GeV$/n, while we use primary/primary ratios to fix the C/O and N/O\footnote{Note that N 
    = $^{14}$N +  $^{15}$N is a combination of primary and secondary nuclides.} injection ratios.

  As in \cite{Evoli:2008dv}, we accomplish this by sampling, for each pair ($D_0/z_t$, $\delta$), the parameter space (C/O, N/O) 
  and computing the $\chi^{2}$ of our predictions for the C/O and N/O modulated ratios against experimental data over the energy 
  range of our interest. For the set of parameters that minimizes this $\chi^{2}$, we then compute the $\chi^{2}$ 
  (which we call $\chi^{2}_{\rm B/C}$) of our predictions for the B/C ratio against data.
  By iterating this procedure for several values of the pair ($D_0/z_t$, $\delta$), we sample the whole parameter space of our interest. 
  Minimization of $\chi^{2}_{\rm B/C}$ leads to the best fit values for ($D_0/z_t, \delta$) and the appropriate confidence regions.
  
  \subsubsection{Antiprotons}

  The construction of a statistically meaningful variable for the $\bar{p}/p$ ratio is rather simpler than for the B/C. Indeed, if we 
  neglect the systematic uncertainties associated to the production and interaction cross sections, the propagation of secondary
  antiprotons depends essentially on $D_{0}/z_{t}$, $\delta$ and the source abundance ratio He/p. This last unknown can be easily fixed 
  by looking at the measured spectrum of He at Earth, which is relatively well known. Therefore, we construct a $\chi^{2}_{\pbar/p}$ by 
  comparing our predicted $\pbar/p$ spectrum for different values of ($D_{0}/z_{t},\delta$) to experimental observations.
  
  \subsubsection{Joint comparison} \label{subsec:joint}
  
  Since we have two different data-sets for the same physical framework, it is important to find links for joint optmization of the combined analysis. 
  The two data-sets being uncorrelated, it is possible to define a joint $\chi^{2}$ knowing the single results of the previous analysis 
  \begin{equation}
    \chi^2 = \frac{1}{2} \left( \chi^2_{B/C} + \chi^2_{\bar{p}/p} \right)\;
    \label{eq:jointchi2}
  \end{equation} 
  and minimize it with respect to ($D_0/z_t,\delta$).
  
  The complementary nature of the two data sets will be demonstrated in section \ref{subsec:results} in which the 
  overlapping of the separated analysis is evident.
  
  \subsection{Experimental Data}
  So far the best B/C measurements above $1~\GeV$/n have been provided by the HEAO-3 \cite{HEAO-3} and CRN \cite{CRN} experiments
  in the range $1 < E_k < 30~\GeV$/n and $ 70~\GeV\mathrm{/n}< E_k \simleq 1.1~\TeV$/n. 
  Recently, the CREAM \cite{CREAM} experiment has released data \cite{Ahn:2008my} improving significantly the available statistics 
  at high energy. C/O and N/O data are taken from the same experiments as well. 
  
  For antiprotons we use experimental data released by BESS for the periods 1995-97 \cite{Orito:1999re} and 1998 \cite{Maeno:2000qx} 
  in the energy interval $1-4~\GeV$, and by CAPRICE (1998) \cite{Boezio:2001ac} in the range $3-49~\GeV$. Recently also the PAMELA 
  experiment has released $\bar{p}/p$ data in the energy range $1 - 100~\GeV$ \cite{PAMELA}. We include them in our analysis.

  Solar modulation has been demonstrated to have important effects in the determination of the $\pbar/p$ ratio at low energy. 
  We account for modulation in the force free approximation. For each data set we use the solar potential of the year when data 
  were taken. 

  \subsection{Results}
  \label{subsec:results}
  
  \begin{figure*}[tbp]
    \centering
    \includegraphics[width=4in]{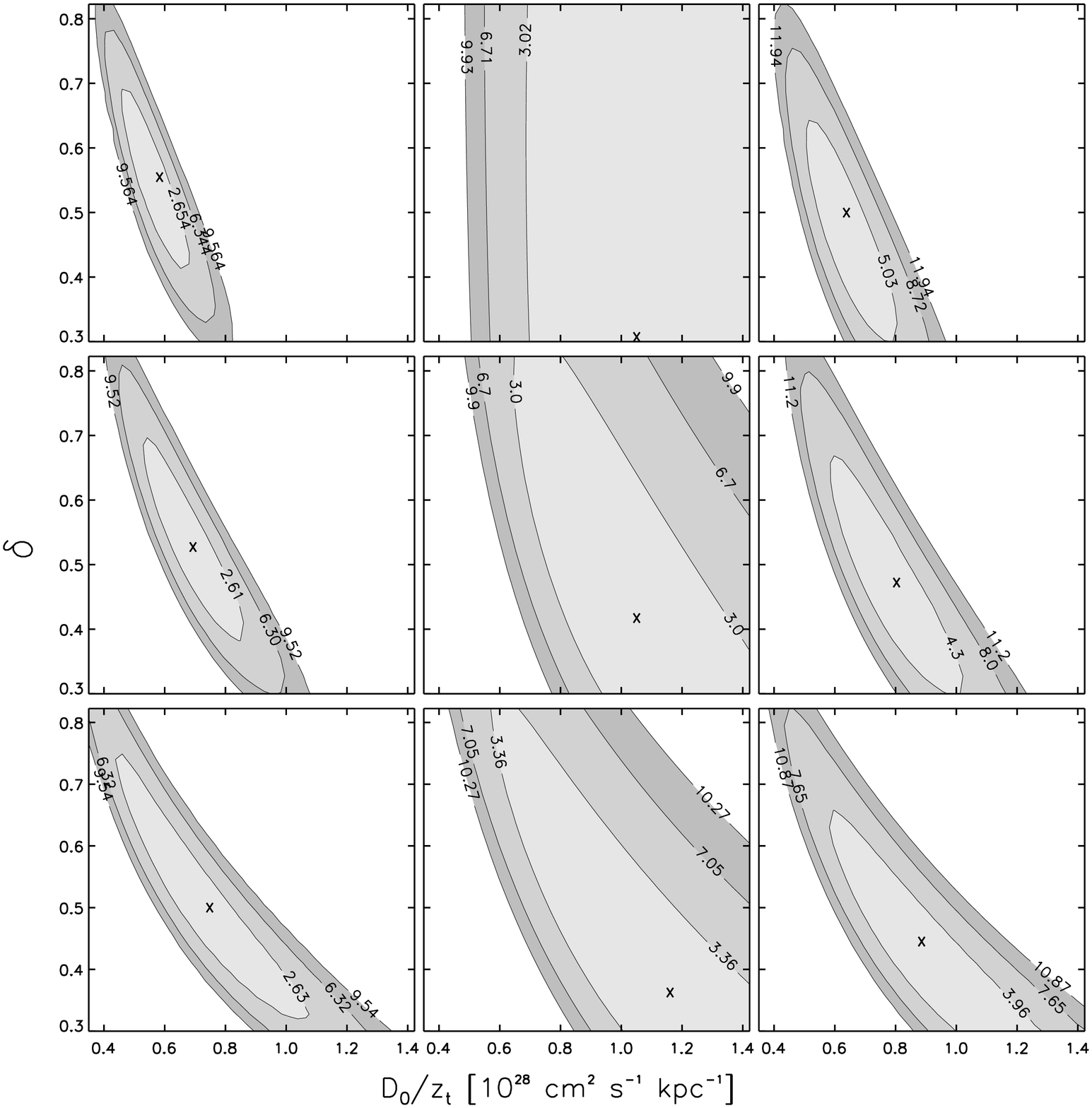}
    \caption{1, 2 and 3$\sigma$ CL regions for different data sets. The left-hand plots represent the CL of $\chi^{2}_{\rm B/C}$, 
      the central plots show the same CL for $\chi^{2}_{\pbar/p}$, while the right-hand side plot is the result of the joint analysis.
      The top panel is obtained for $E_{\rm min} = 1$~GeV/n and with all data before CREAM and PAMELA, the central one for $E_{\rm min} = 1$~GeV/n
      and all data while the bottom panel is for $E_{\rm min} = 5$~GeV/n and all data.}
    \label{fig:combo}
  \end{figure*}
  
  \begin{table*}[ht]
    \caption{Results of the statistical analysis described in Sec.~\ref{sec:analysis}.}
    \label{tab:best_fits}
    \centering
    \begin{tabular}{|c|c|c|c|c|c|c|c|c|c|}
      \hline
      & \multicolumn{3}{|c|}{Min~$\chi^2 ({B/C})$} & \multicolumn{3}{|c|}{Min~$\chi^2({\pbar/p})$} & \multicolumn{3}{|c|}{Joint analysis} \\
      \hline
      Emin  & $D_0/z_t$ & $\delta$ & $\chi^2$ & $D_0/z_t$ & $\delta$ & $\chi^2$ & $D_0/z_t$ & $\delta$ & $\chi^2$ \\
      \hline 
      
       1 & 0.68 & 0.52 & 0.39 & 1.04 & 0.41 & 0.84 & 0.76 & 0.47 & 1.94 \\
       5 & 0.74 & 0.49 & 0.33 & 1.15 & 0.36 & 1.06 & 0.87 & 0.44 & 1.66 \\
      10 & 0.82 & 0.47 & 0.22 & 0.82 & 0.52 & 1.22 & 0.87 & 0.44 & 0.88 \\
      
      \hline
    \end{tabular}
  \end{table*}
  
  In Fig.~\ref{fig:combo} we show our results for the separated and joint analysis of the B/C and $\pbar/p$ data.
  The left-hand plots represent the 1, 2 and 3$\sigma$ contour levels of the $\chi^{2}_{\rm B/C}$, in the $(D_{0}/z_{t},\delta)$ space, 
  the central plots show the same contour levels for $\chi^{2}_{\pbar/p}$, while the right-hand side plot is the result of the joint 
  analysis described 
  in section \ref{subsec:joint}\footnote{Our $\chi^{2}$ variables are always understood to be ``reduced'' $\chi^{2}$.}. 
  From the panels in the first column the impact of CREAM results on our knowledge of the propagation parameters is evident: 
  CREAM high energy data favour a smaller value of $\delta$ with respect to previous ones.  
  Also PAMELA $\pbar/p$ data are sensitive to our propagation parameters, in contrast with pre-PAMELA $\pbar/p$ data (see the upper 
  row panels in Fig.~\ref{fig:combo}). 
  Finally, it is clear from the lower row panels as for $E_{\rm min} >  5~\GeV$ there is an almost complete concordance between the CL 
  regions constrained by high energy B/C and  $\pbar/p$ data. Indeed, even the 1$\sigma$ regions do have a significant overlapping and 
  the minimum joint $\chi^{2}$ is 1.2 for $E_{\rm min} = 6~\GeV$/n. 
  The joint analysis indicates a best-fit value of $\delta \simleq 0.5$, possibly favouring a Kraichnan power spectrum for the turbulent 
  galactic magnetic field. While the best fit for $\delta$ seems not to be strongly dependent on $E_{\rm min}$, we find that the best value 
  of $D_{0}/z_{t}$ tends to be larger when a larger $E_{\rm min}$ is considered. This could indicate, in agreement with na\"ive expectations, 
  that the scale of vertical diffusion $z_{t}$ is smaller at higher energies.

\begin{figure*}[tbp]
    \centering
    \includegraphics[width=2in]{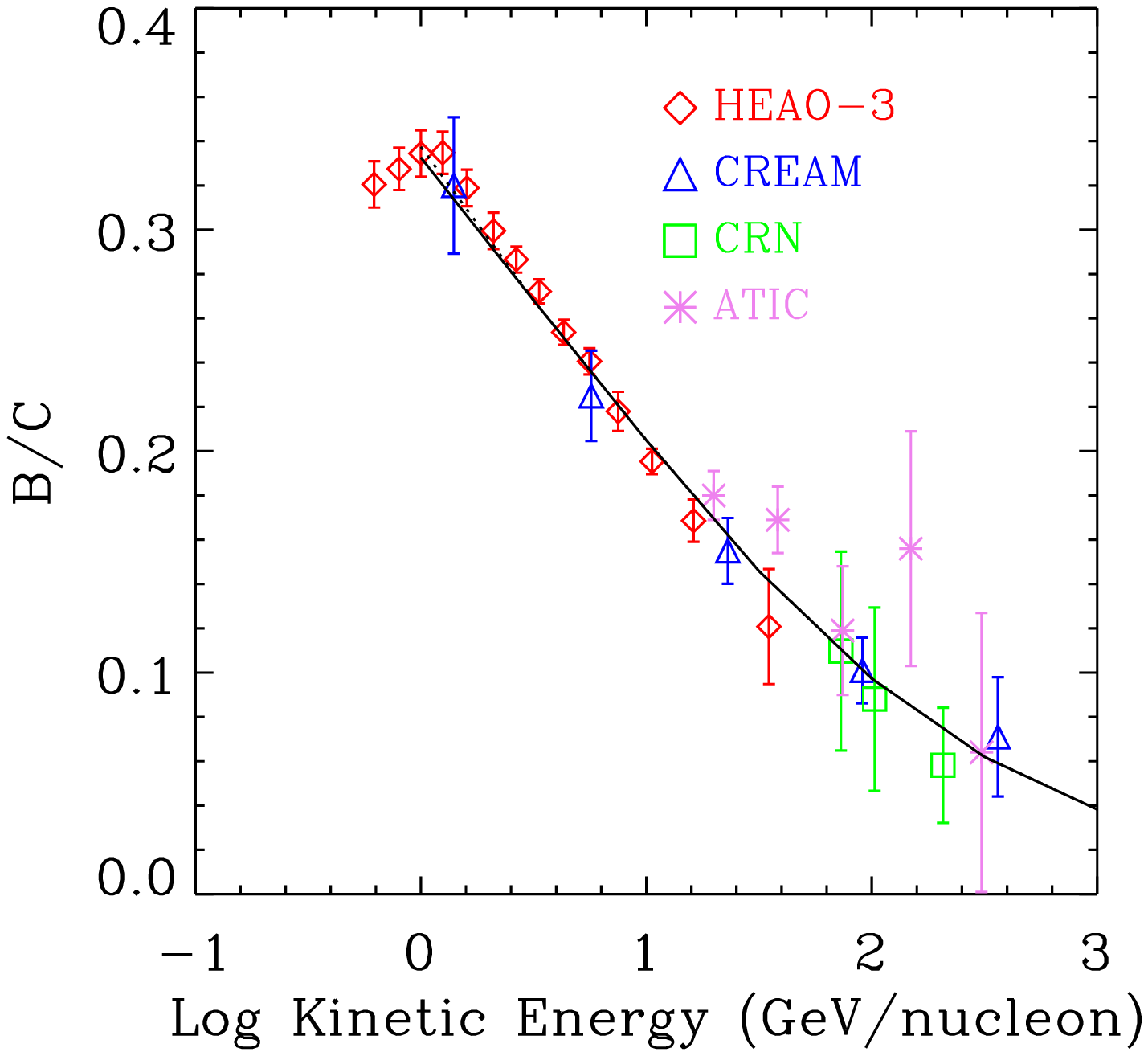}
    \includegraphics[width=2in]{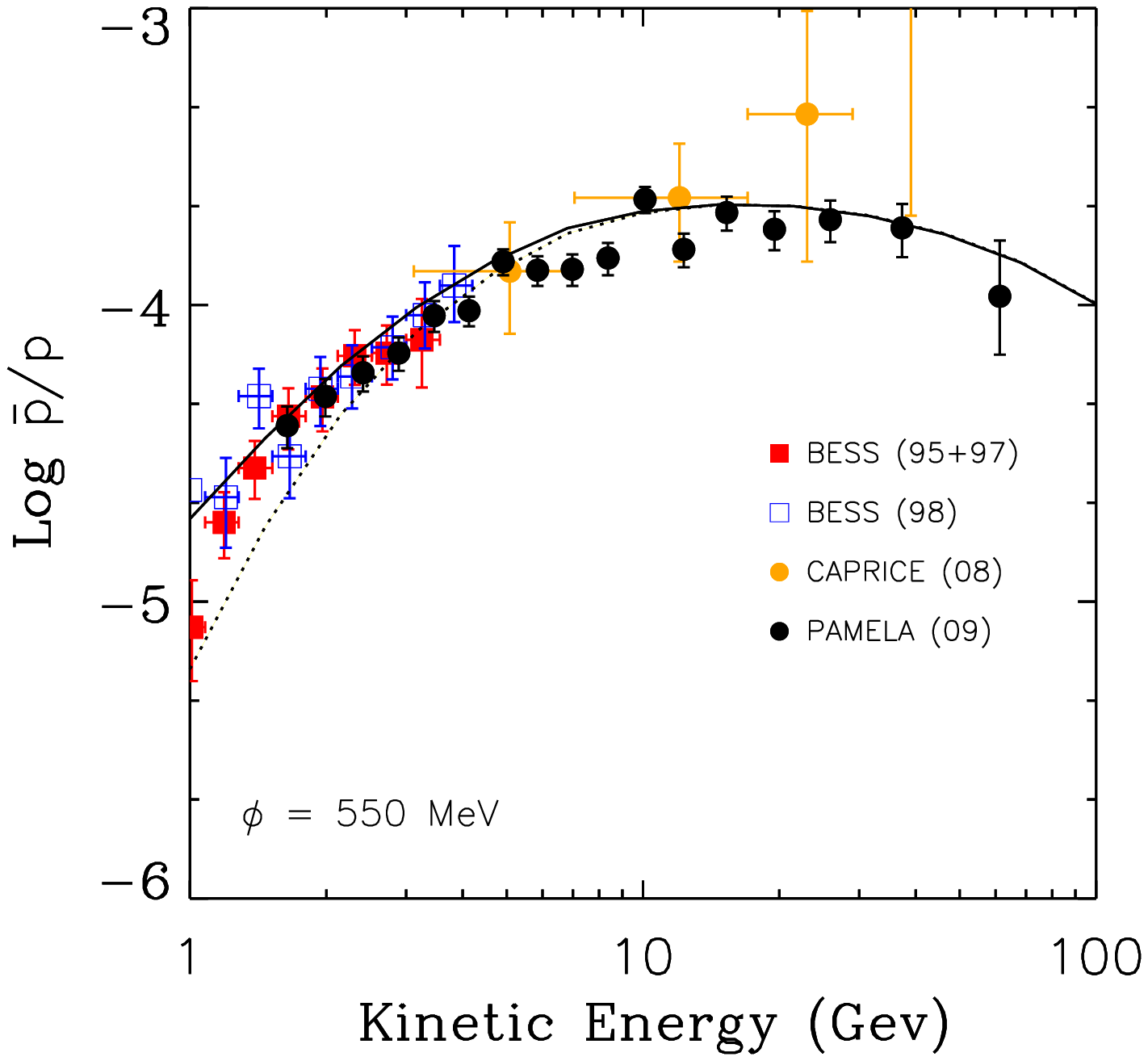}
    \caption{Comparison of experimental (top of the atmosphere) data for B/C (left panel) and ${\bar p}/p$ ratios with our joint analysis best-fit 
      model (second row in Tab.~\ref{tab:best_fits}). Dotted lines refer to LIS ratios, accounting for solar modulation with potential $\phi = 550~{\rm MV}$.}
    \label{wide_fig}
  \end{figure*}

  The concordance between nuclear and antiproton data is also evident from Fig.~\ref{wide_fig} where the predictions of the combined analysis 
  best-fit model are compared with B/C and $\pbar/p$ experimental data. 

  Finally, in Tab.~\ref{tab:best_fits} we recap the findings of our minimization strategy. Best fit values for ($D_{0}/z_{t},\delta$) 
  are obtained considering first the two data sets separetely and then jointly.

  Our analysis shows clearly that the two data sets are statistically compatible, within our model. 
  
  \section{Conclusions}
  
  We performed a statistical analysis to constrain the CR propagation parameters in a plain diffusion scenario numerically implemented in DRAGON. 
  Taking advantage of the new CREAM and PAMELA high-energy data we performed a combined analysis of B/C and ${\bar p}/p$ data in several energy ranges.   
  This approach allowed us to test the (weak) dependence of our results on poorly know low-energy physics. 
  We showed that above few GeV/n the whole data sets considered here are consistently reproduced by a PD model. Our findings favour a Kraichnan type 
  ($\delta \simeq 0.5$) dependence of the diffusion coefficient on energy.

\end{document}